# Evaluation of Burst Loss Rate of an Optical Burst Switching (OBS) Network with Wavelength Conversion Capability

Md. Shamim Reza, Md. Maruf Hossain and Satya Prasad Majumder

**Abstract** – This paper presents a new analytical model for calculating burst loss rate (BLR) in a slotted optical burst switched network. The analytical result leads to a framework which provides guidelines for optical burst switched networks. Wavelength converter is used for burst contention resolution. The effect of several design parameters such as burst arrival probability, wavelength conversion capability, number of slots per burst and number of wavelengths is incorporated on the above performance measure. We also extend the analytical result of BLR for different types of service classes where each service class has a reserved number of wavelengths in a network with fixed number of wavelengths. We also introduce an algorithm to calculate the resultant number of wavelength for each service classes depending on the various scenarios.

————————— ◆ —————————

## 1 INTRODUCTION

IN future telecommunication networks, traffic with different performance requirements will be merged in the same physical layer, and will require new, adaptable network architectures [1]. This has been forcing the backbone network to evolve quickly from an electronic network to an optical network, which requires high-speed and high performance optical-switching technologies [2]. In order to utilize efficiently the amount of raw bandwidth in wavelength division multiplexing (WDM) networks, all-optical transport method must be developed to avoid electronic buffering while handling bursty traffic [3]. Several different technologies have been proposed for the transfer of data over dense wavelength-division multiplexing. Wavelength-routed or optical-circuit-switching (OCS) networks have already been deployed. However, being a form of circuit-switching network, OCS is not optimally bandwidth-efficient for IP traffic. Optical packet switching (OPS) performs packet switching in which a packet is sent along with its header. While the header is being processed by an intermediate node, either all-optically or electronically (after an O/E conversion), the packet is buffered at the node in the optical domain. However, high-speed optical logic, optical-memory technology, and synchronization requirements are major problems with this approach [4]. The optical burst switching (OBS) that was proposed in [4] and [5] has been increasingly receiving attention as a potentially bandwidth efficient approach for supporting an all-optical Internet.

Optical burst switching (OBS) is a new wave length division multiplexing technology that combines the advantages of both wavelength-routed (WR) networks and optical packet switching (OPS) networks [6], [7]. As in WR networks, there is no need for buffering and electronic processing for data at intermediate nodes. At the same time, OBS increases the network utilization by reserving the optical channel for a limited time period.

An OBS network is composed of edge and core nodes. IP packets are assembled into a burst at the ingress edge node, and the burst is routed over a buffer-less core network [2]. When two or more bursts are destined for the same output port at the same time, contention occurs; there are many contention resolution schemes that may be used to resolve the contention. The primary contention-resolution schemes are optical buffering, wavelength conversion, deflection routing and burst segmentation.

Optical buffer is usually composed of fiber delay lines (FDLs). While the contending bursts are transmitted in FDL with a fixed length the transmission delay caused by light wave transmission delay in optical fibre can be used as the optical buffering time. Once an optical burst has entered into the optical fibre, it must come out from the other end after a certain time [8].

In wavelength conversion, if two bursts on the same wavelength are destined to go out of the same port at the same time, then one burst can be shifted to a different wavelength [9].

In deflection routing, one of the two bursts will be routed to the correct output port (primary) and the other to any available alternate output port (secondary). The deflected packets may end up following a longer path to the destination, leading to higher end-to-end delay, and packets may also arrive at the destination out of order [10].

In burst segmentation [11], the burst is divided into basic transport units called segments. Each of these segments may consist of a single IP packet or multiple IP packets, with each segment defining the possible partitioning points of a burst when the burst experiences contention in the optical network. All segments in a burst are initially transmitted as a single burst unit. However, when contention occurs, only those segments of a given burst that overlap with segments of another burst will be dropped. If switching time is not negligible, then additional segments may be lost when the output port is switched from one burst to another. There are two approaches for dropping burst segments when contention occurs between bursts. The first approach, tail dropping, is to drop the tail of the original

————————————————

- Md. Shamim Reza is with the Dept. of Electrical and Electronic Engineering, Bangladesh University of Engineering and Technology, Dhaka-1000, Bangladesh.
- Md. Maruf Hossain is with the Dept. of Electrical and Electronic Engineering, American International University of Bangladesh, Dhaka, Bangladesh.
- Satya Prasad Majumder is with the Dept. of Electrical and Electronic Engineering, Bangladesh University of Engineering and Technology, Dhaka-1000, Bangladesh.

burst, and the second approach, head dropping, is to drop the head of the contending burst [11]. A combination of contention-resolution techniques may be used to provide high throughput, low delay, and low packet-loss probability.

The main objective of this paper is to present an analytical model that captures the essential features of JIT-based OBS (OBS-JIT) networks, and gives stronger insight into factors that affect the burst loss rate (BLR). We start by briefly describing the background of OBS in Section 2. In Section 3, we briefly describe previous models for calculating the burst blocking probability in [6] and then we develop a model for calculating the burst loss rate. In section 4, we extend the analytical model of BLR for different types of service classes. Model results are presented in section 5. Finally we give our conclusion in section 6.

## 2 BASIC OF OBS

In an OBS network, the basic switching entity is a burst. A burst is a train of packets moving together from one ingress node to one egress node and switched together at intermediate nodes [7]. An ingress node assembles the internet protocol packets, which are coming from the local access networks and destined to the same egress node, into large burst [6]. An optical burst consists of two parts, a header and a data burst. The header is called the control burst (CB) and is transmitted separately from the data, which is called data burst (DB). Data bursts are disassembled at the egress router into individual packets. These ingress and egress nodes, where aggregation and distribution of data from and to access networks take place, are called as edge nodes. The nodes of the core network, through which the aggregated data is transmitted, are referred to as core nodes. This process of aggregating several packets into a burst is called burst assembly. During burst assembly, data is buffered at the edge node where electronic RAM is easily available. Before each data burst is transmitted, a control burst is transmitted over a separate control channel. The control burst contains an information about the sender, receiver and transmission wavelength of the corresponding burst. This control burst, which is sent ahead of the data burst, undergoes Optical/Electronic/Optical (O/E/O) conversion and is processed electronically at each intermediate node, to configure the switch for the data burst that is to arrive later. This process is called as burst reservation.

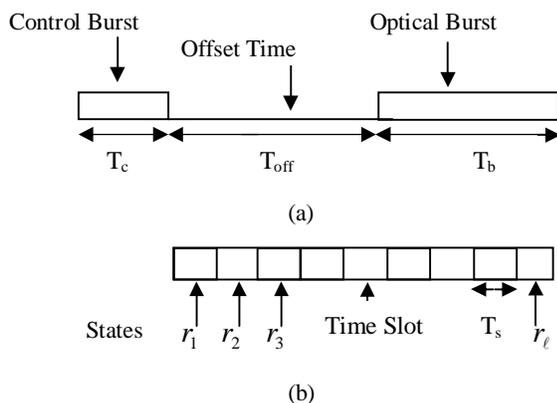

Fig. 1. (a) Transmission of an optical burst (b) Slotted timing diagram.

The offset time, which is the difference between the end time of control burst and staring time of data burst at the ingress node, is such that the data burst is switched all-optically at any intermediate node, without incurring any delay (fig 1a). The value of offset time is chosen to be greater than or equal to the total processing time delay encountered by the CB [7]. Hence no optical RAM or buffers are needed at any intermediate node. An intermediate node gets many control bursts and it needs to configure the switches for all the corresponding data bursts and decide channels (wavelengths) on which to schedule each data burst. This process is called as burst scheduling.

### 2.1 Burst Aggregation

One of the main functions of an OBS user is to collect upper layer traffic, sort it based on destination addresses, and aggregate it into variable-size bursts. The exact algorithm for creating the bursts can greatly impact the overall network operation because it allows the network designers to control the burst characteristics and therefore shape the burst arrival traffic.

The burst assembly algorithm has to consider the following parameters: a preset timer, and maximum and minimum burst lengths. The timer is used by the user in order to determine when exactly to assemble a new burst. The maximum and minimum burst parameters shape the size of the bursts. This is necessary since long bursts may hold resources for long times and cause higher burst losses, while short bursts may give rise to too many control packets. The burst aggregation algorithm may use bit-padding if there is not enough data to assemble a minimum size burst. The most common burst-assembly approaches are timer based and threshold based. In a timer-based burst-assembly approach, a burst is created and sent into the optical network when the time-out event is triggered. In a threshold-based approach, a limit is placed on the number of packets contained in each burst. A more efficient assembly scheme can be achieved by combining the timer based and threshold-based approaches.

### 2.2 Signaling Protocols

Several signaling protocols have been proposed for OBS [4], [5], [7]. OBS is based on one-way reservation protocol, such as Just-Enough-Time (JET), Just-In-Time (JIT) and Tell-And-Go (TAG), in which a data burst follows a corresponding control packet without waiting for an acknowledgement. Among these, two protocols are widely studied in the literature: Just-In-Time (JIT) [7] and Just-Enough-Time (JET) protocol [4], [7]. Other protocols can be considered a variation of one of these two [7].

In this section, we briefly describe the operation of the JIT and JET protocols. The JIT protocol works as follows. Upon the arrival of a CB to a core OBS node, a wavelength channel is immediately reserved if available. If no wavelength is available the request is blocked and the corresponding DB is dropped. When the wavelength is successfully reserved, it remains reserved until the data burst transmission has finished. The only information that needs to be kept in the network nodes is whether the wavelength is currently reserved or not. The control packet is not aware of the burst length and reserves the relevant link bandwidth (if available) for the entire burst as soon as it arrives at the switch. The JET protocol, on the other hand, is reserve-a-fixed duration scheme that reserves resources exactly for the transmission time of the data burst. In JET, when a CB arrives at a core node, a wavelength channel scheduling algorithm is invoked to find a suitable wavelength channel on the outgoing link for the corresponding DB. The wavelength channel is reserved for a duration equal to the burst length starting from the arrival time of the DB. The information required by the scheduler such as the data burst arrival time and its duration are obtained from the control burst [7].

### 2.3 Wavelength Allocation: With or Without Conversion

In an OBS network with no wavelength converters, the entire path from source to destination is constrained to use a single wavelength. The other possibility is an OBS network with a wavelength conversion capability at each OBS node. In this case, if two bursts contend for the same wavelength on the same output port, the OBS node may optically convert one of the signals from an incoming wavelength to a different outgoing wavelength. In addition, the conversion capability at an OBS node can be classified further as full or sparse. In the former case, there is one converter per each wavelength, whereas in the latter case the number of converters is less than the total number of wavelengths. Wavelength conversion is a desirable characteristic in an OBS network as it reduces the burst loss probability. However, it may not necessarily be a practical assumption since all optical converters are still an expensive technology.

### 2.4 Scheduling of Resources: Reservation and Release

Upon receipt of the control packets sent from the OBS users, the OBS nodes schedule their resources based on the included information. The proposed OBS architectures differ in their resource (wavelength) reservation and release schemes. Baldine et al. [12] classified these schemes based on the amount of time a burst occupies a path inside the switching fabric of an OBS node. In explicit setup, a wavelength is reserved, and the optical cross connect is configured immediately upon processing of the control packet. In estimated setup, the OBS node delays reservation and configuration until the actual burst arrives. The allocated resources can be released after the burst has come through using either explicit release or estimated release. In explicit release, the source sends an explicit trailing control packet to signify the end of a burst transmission. In estimated release, an OBS node knows exactly the end of the burst transmission from the burst length, and therefore can calculate when to release the occupied resources. Based on this classification, the following four possibilities exist: explicit setup/explicit release, explicit setup/estimated release, estimated setup/explicit release, and estimated setup/estimated release.

## 3 EVALUATION OF BURST LOSS RATE (BLR)

Let us consider a slotted burst switch with one input and output fiber, where the fiber provides $w$ wavelengths by using wavelength-division multiplexing. When bursts arrive on the input wavelengths to the switch in a given time slot, the control bursts are processed electronically. Based on the desired information extracted from the control packets, the control module decides which wavelengths the bursts are switched to and configure the switch fabric accordingly.

The switch has wavelength converters (WC) with $\rho$ conversion capability at each output. If $\rho = 0$, then the network has no conversion capability, whereas if $\rho = 1$, then the network has a full conversion capability, it can be assumed that, at any node, all wavelengths are available in a pool. The effect of the switching time is ignored.

In this paper we consider JIT-based OBS protocol but a little modification of resource reservation in [6]. The resource will be reserved just after the arrival of control burst and optical burst arrives at the OBS node under consideration after an offset time $T_{off}$ from the arrival of the control burst, which takes care of the processing and configuration times of the control burst and OXC (optical cross connect) fabric, respectively. The total time T spent from the transmission of the control packet until the end of the optical burst is $T = T_c + T_{off} + T_b$, where $T_c$ and $T_b$ are the control burst and optical data burst time durations, respectively.

We divide the resource reservation period $(T_{off} + T_b)$ into small time slots, each of duration $T_s$ called slot time (fig 1). The total number of slots $\ell$ is calculated as

$$\ell = \frac{T_{off} + T_b}{T_s}$$

Where, we assume that $\ell$ is an integer and $T_s$ is a multiple of the bit duration and will be held fixed. It should be emphasized that during a time slot $T_s$ s, the node would get enough information about the selected wavelength. Here we only need to reserve the wavelength for a time duration $T_{off} + T_b$.

From the state-diagram described in [6], we assume that initially, an OBS node is in initial state $m$. If there is an arrival with probability $A$, the OBS node will enter the following state $r_1$ and starts processing the control burst. On the other hand, if there is no arrival with probability $1 - A$, the OBS node will remain as is.

Let $r_1, r_2, \ldots, r_\ell$ are the 1-$\lambda$ states (fig 1). If an OBS node is in state $m$ and there is an arrival, it will reserve one of the available wavelengths and enter state $r_1$. The node in $r_1$ state is serving slot 1 of the first burst. If it is in state $r_1$ and there are no arrivals after $T_s$, it enters state $r_2$. The node in $r_2$ state is serving slot 2 of the first burst. On the other hand, if it is in state $r_1$ and there is an arrival that needs to use the same wavelength, the burst will be blocked with probability $A/w$ and the node will again enter state $r_2$. However, if the node is in state $r_1$ and there is an arrival that needs to use another wavelength (an event that occurs with probability $A(w-1)/w$, the node will serve both bursts and enters state $r_{21}$. The node in $r_{21}$ state is serving slot 2 of the first burst, slot 1 of the second burst [6].

Let us consider an OBS network with $w$ number of wavelengths and the network is equipped with $u$ number of wavelength converters. Only $u$ number of wavelengths among $w$ wavelengths can be converted but the remaining $w-u$ wavelengths cannot be converted. So we can define the wavelength conversion capability by the following equation:

$$\rho = \frac{u}{w}, \qquad 0 \le \rho \le 1$$

When an arriving burst is to be served with a specific wavelength, this wavelength is removed from the pool until after the service is complete. If another arriving burst is to be served with a wavelength not available in the pool, it will be converted to another one from the pool. This latter wavelength is then removed, and $u$ is decreased by one. Blocking occurs whenever the pool is empty, or a used wavelength is needed while $u = 0$.

Assume that $w < \ell$, and consider an $n$-$\lambda$ state $r_{i_n i_{n-1} i_{n-2} \ldots i_2 i_1}$ where $n \in \{1, 2, \ldots \ell \wedge w\}$ and $i_1, i_2, i_3, \ldots i_n \in \{1, 2, 3 \ldots \ell\}$ with $i_n > i_{n-1} > i_{n-2} > \ldots > i_1$, here $\ell \wedge w = \min(\ell, w)$. The node in this state is serving slot $i_n$ of the first burst, slot $i_{n-1}$ of the second burst and so one. After $T_s$ s, if there is an arrival that needs to be served.

Let us consider

$$r_{i_n i_{n-1} i_{n-2} \ldots i_2 i_1} = e_n$$

Expression for $e_n$ is available in [6], we get for any $k \in \{1, 2, \ldots, \ell \wedge w\}$

$$e_k = \frac{\prod_{i=0}^{k-1} \frac{w - i(1-\rho)}{w\frac{1-A}{A} + i(1-\rho)}}{1 + \sum_{n=1}^{\ell \wedge w} \binom{\ell}{n} \prod_{i=0}^{n-1} \frac{w - i(1-\rho)}{w\frac{1-A}{A} + i(1-\rho)}}$$

Thus if $n \neq w$, the blocking probability [6] for this node is given by

$$P_b(n) = \frac{A(\ell - 1)}{w\ell}(1-\rho) n \binom{\ell}{n} e_n$$

If $n = w$, however, blocking probability [6] is given by

$$P_b(w) = \frac{A(\ell-1)}{w\ell}(1-\rho) w \binom{\ell}{w} e_w + A\rho \binom{\ell-1}{w} e_w$$

We denote the burst arrival probability on an input wavelength in a given time slot as $A$ ($0 \leq A \leq 1$) and assume that $A$ is independent on burst arrivals in other wavelengths and burst arrivals on in previous time slots.

Let $A_k$ ($0 \leq A_k \leq 1$) be the probability for $k$ ($0 \leq k \leq w$) arrivals to the output fiber on a given time slot. $A_k$ is then distributed according to a Binomial process $P_A(k/w)$ [13].

$$A_k = P_A(k/w) = \binom{w}{k}(A)^k (1-A)^{w-k} \quad (1)$$

The average number of burst arrivals in a time slot is $E[A_k] = Aw$. If two control bursts are to reserve the same wavelength at a given core node for two different bursts, then only one burst will be offered to this wavelength. The other will be blocked and lost unless there is an available wavelength converter. So we obtain the average burst loss rate which is given in equation 2.

$$BLR = \frac{1}{Aw} \sum_{k=1}^{\ell \wedge w} A_k \cdot k \cdot P_b(k) \quad (2)$$

## 4 BLR DIFFERENTIATION OF DIFFERENT TYPE OF SERVICE CLASSES

Let the network has $d$ service classes, ranging from service class 0 to service class $d-1$. Let $BLR_i$ ($0 \leq i \leq d-1$) be the burst loss rate for class $i$ traffic at the output fibre, and let $S_i$ be the relative share of class $i$ traffic.

In order to isolate the service classes, we introduce the parameter $L_i$ ($0 \leq L_i \leq N$), which is the number of wavelengths reserved for class $i$ traffic in the case of contention in a time slot. Let $j_i$ denote the number of class $i$ bursts that arrive in a time slot. If the total number of burst arrival to the input fibre in a time slot is $k$, we must have $j_0 + j_1 + \ldots + j_{d-1} = k$. Regarding service class $i$, if $j_i \leq L_i$ then $L_i - j_i(1 - P_b(j_i))$, wavelengths in the considered time slot that are not utilized by the class $i$ bursts are denoted as free wavelengths, where $P_b(j_i)$ denotes the burst blocking probability for $j_i$ number of bursts. On the other hand if, $j_i \geq L_i$, there are more bursts than reserved wavelengths available for service class $i$. In this case, maximum $L_i$ bursts will be transmitted, however, the minimum $j_i - L_i$ bursts that do not get a wavelength among the $L_i$ reserved wavelengths are denoted as overflow bursts with full wavelength conversion capability ($\rho = 1$) and $w \geq \ell$, where $w$ is the resultant number of wavelength of service class $i$. But if $\rho < 1$ then the number of overflow bursts will be greater than $j_i - L_i$. Overflow bursts will attempt to seize free wavelengths from other service classes, if available, in order to ensure that all wavelengths are utilized. In any case, we do not leave any wavelengths idle when contention occurs.

Here we consider d=2 i.e. two different service classes have been considered. So the average burst loss rate ($BLR_i$) for class $i$ traffic is given in equation (3), where $A_k$ is given in equation (1), $M(.)$ is given in equation (4) which is the multinomial distribution and $L_{lost}$ is the number of lost bursts for service class $i$. $L_{lost}$ can be calculated from the following algorithm:

```
IF ( j_0 < L_0 ) THEN DO
    IF ( j_1 < L_1 )
        L_{0_new} = L_0 + [ L_1 - j_1(1 - P_b(j_1)) ]
        L_{1_new} = L_1 + [ L_0 - j_0(1 - P_b(j_0)) ]
    ELSE
        L_{0_new} = L_0
        L_{1_new} = L_1 + [ L_0 - j_0(1 - P_b(j_0)) ]
    END
ELSE
    L_{0_new} = L_0 + [ L_1 - j_1(1 - P_b(j_1)) ]
    L_{1_new} = L_1
END
L_{i_new} = round ( L_{i_new} )
IF    ( j_i < L_{i_new} )
    L_{lost} = j_i . P_b(j_i)
ELSE
    L_{lost} = ( j_i - L_{i_new} ) + L_{i_new} . P_b(L_{i_new})
END
```

Here, $L_{0\_new}$ and $L_{1\_new}$ approximately denote the number of wavelengths for service class 0 and service class 1 respectively and we consider that $w$ is the resultant wavelengths for any service class and the number of burst arrival is $n$ at any instant of time. For service class 0, $w = L_{0\_new}$ and for service class 1, $w = L_{1\_new}$.

## 5  MODEL RESULTS

The model results have been shown below. Fig. 2 & 3 show the effect of burst arrival probability on BLR. BLR increases with the increasing of burst arrival probability. It also shows that for constant burst arrival probability, burst loss rate decreases with the increasing of wavelength conversion capability and number of channel wavelengths. The effect of wavelength conversion capability on the burst loss rate is shown in fig. 4 & 5. If the number of wavelength converter is fixed then loss increases as the arrival increases and loss decreases as the number of channel wavelength increases. With constant burst arrival probability, network traffic (traffic=A. $\ell$) increases as the number of slots per burst increases, so fig. 6 and 7 show that burst loss rate increases with the increasing of the network traffic. The burst loss rate decreases with the increasing of the number of channel wavelength which is shown in fig. 8 and 9. When the number of channel wavelength is one then the loss rate doesn't depend upon the wavelength conversion capability but burst loss rate depends upon the burst arrival. Fig. 10 shows the effect of burst blocking probability on burst loss rate. The entire burst will be lost at the unity blocking probability.

The presented analytical model for d service classes has been evaluated for a d = 2 service class scenarios. Here only one input and one output fiber is used. We have used rounded $L_{0\_new}$, rounded $L_{1\_new}$, total number of wavelengths $N$ =16, $S_0$ =0.5, $S_1$ = 0.5, $L_1 = N - L_0$ and $\ell$ =100.

We see that BLR for class 0 traffic decreases as $L_0$ increases which are shown in fig 11 and 12 but burst loss rate for class 1 traffic increases as $L_0$ increases which are shown in fig. 13 and 14. This is expected since an increase in $L_0$ means that more wavelengths are reserved for class 0 traffic, and fewer wavelengths are reserved for class 1 traffic. The effects of wavelength conversion capability and burst arrival probability have also been shown in fig. 11-14. Results show that system performance improves with the increasing of wavelength conversion capability but degrades as the network traffic increases. Fig. 15 shows BLR for both the service classes on the same plot. When $L_0$ =8, then the loss rate of both the service classes is same because each of the service class has equal number of reserved wavelengths which is 8.

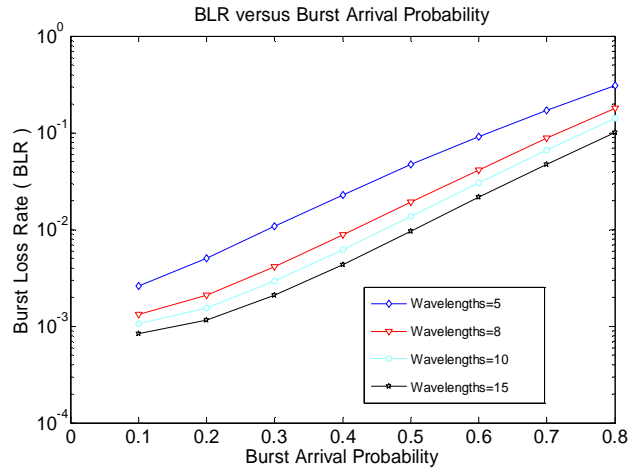

Fig. 2. Burst loss rate versus burst arrival probability with ρ=0.1, $\ell$ =100 and $w < \ell$.

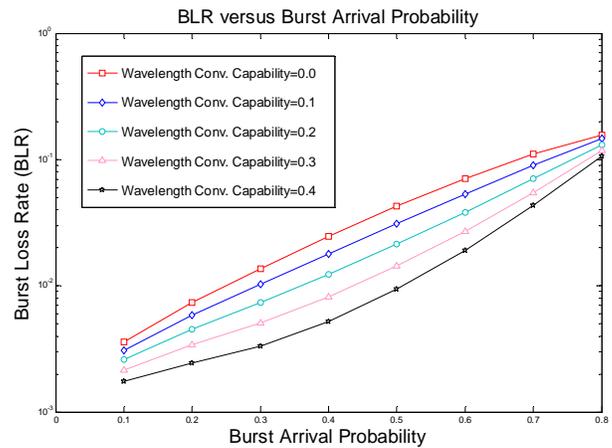

Fig. 3. Burst loss rate versus burst arrival probability with $w$ = 20, $\ell$ = 100 and $w < \ell$.

---

$$BLR_i = \frac{1}{ANS_i} \sum_{k=1}^{\ell \wedge N} A_k \left[ \sum_{j_o=0}^{k} \cdot \sum_{j_1=0}^{k-j_o} \cdots \sum_{j_{d-2}=0}^{k-\sum_{v=0}^{d-3} j_v} \left[ M\left(j_0, \ldots j_{d-2}, k - \sum_{v=0}^{d-2} j_v ; S_0, S_1, \ldots, S_{d-1}\right) \times L_{lost} \right] \right] \quad (3)$$

$$M\left(j_o, \ldots \ldots j_{d-2}, k - \sum_{v=0}^{d-2} j_v ; S_o, S_1, \ldots, S_{d-1}\right) = \binom{k}{j_0 j_1 \ldots j_{d-2} k - \sum_{v=0}^{d-2} j_v} S_o^{j_o} S_1^{j_1} \ldots \ldots S_{d-1}^{j_{d-1}} \quad (4)$$

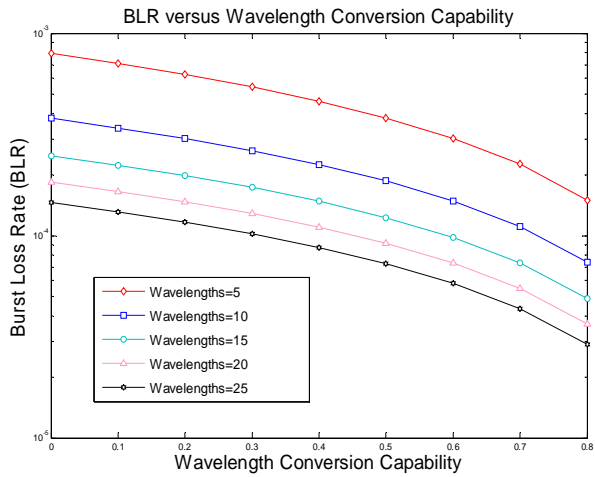

Fig. 4. Burst loss rate versus wavelength conversion capability with $\ell$ =100, A=0.01 and $w < \ell$.

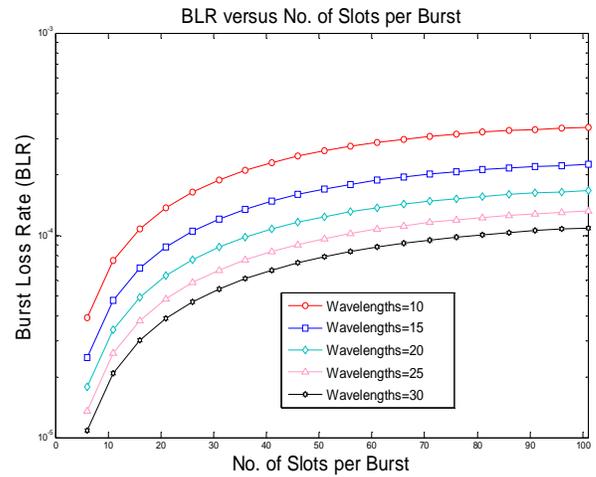

Fig. 7. Burst loss rate versus number of slots per burst with A = 0.01 and ρ = 0.1.

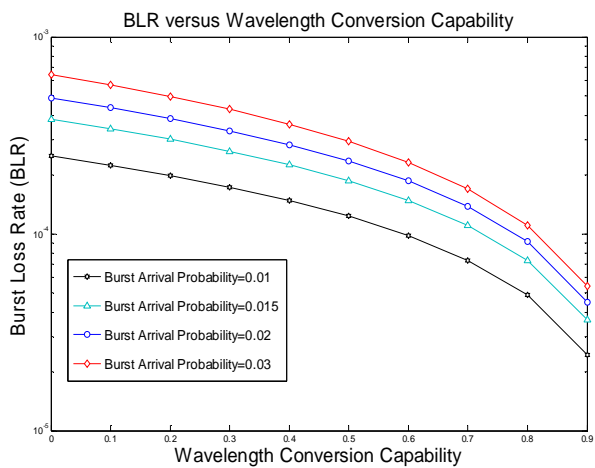

Fig. 5: Burst loss rate versus wavelength conversion capability with $\ell$ =100, $w$ = 15 and $w < \ell$.

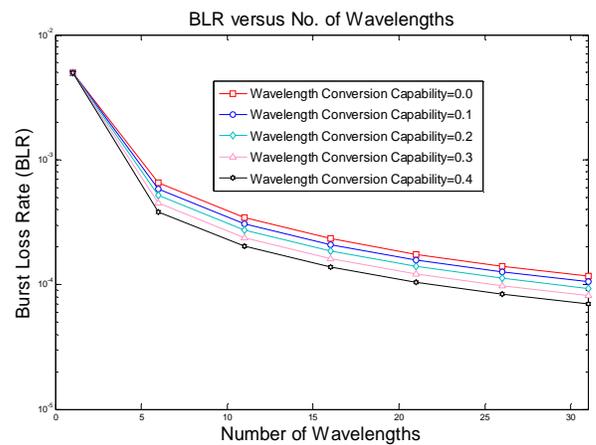

Fig. 8. Burst loss rate versus total number of wavelengths with $\ell$ =100, A=0.01 and $w < \ell$.

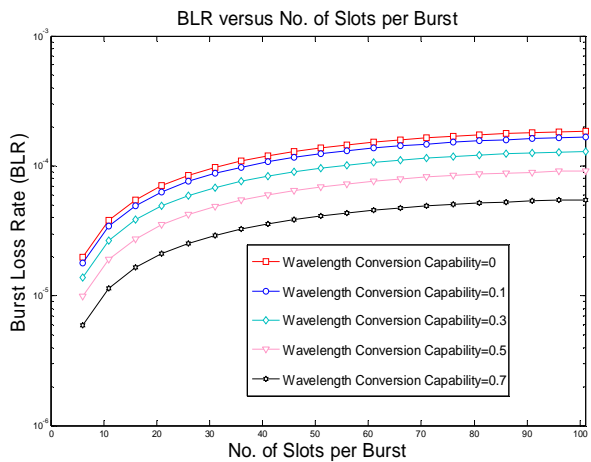

Fig. 6. Burst loss rate versus number of slots per burst with A = 0.01 and $w$ = 20.

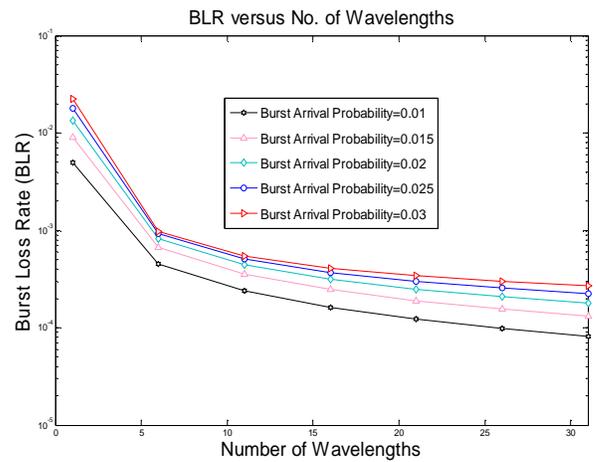

Fig. 9. Burst loss rate versus total number of wavelengths with $\ell$ =100, ρ =0.3 and $w < \ell$.

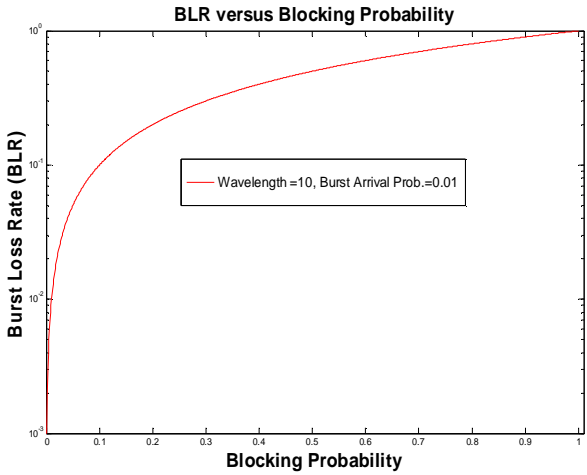

Fig. 10. The effect of burst blocking probability on average burst loss rate. Here $W$ = 10, $w < \ell$ and A=0.01.

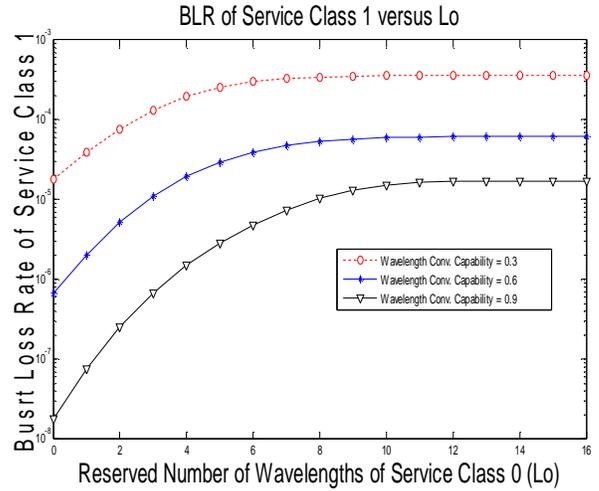

Fig. 13. The effect of reserved wavelengths (Lo) for service class 0 on burst loss rate of the other service class with $A$ = 0.5.

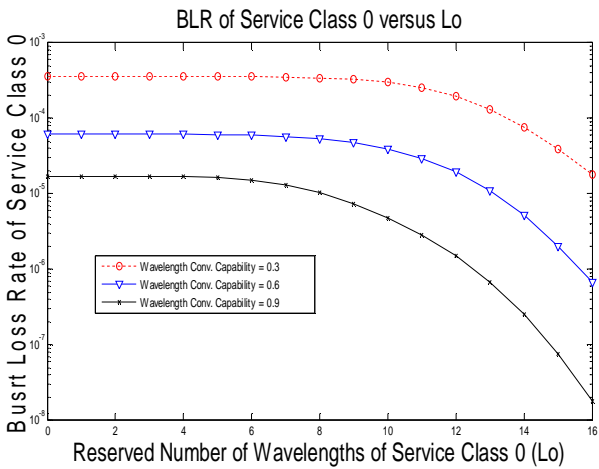

Fig. 11. The effect of reserved wavelengths (Lo) for service class 0 on burst loss rate of the same service class with $A$ = 0.5.

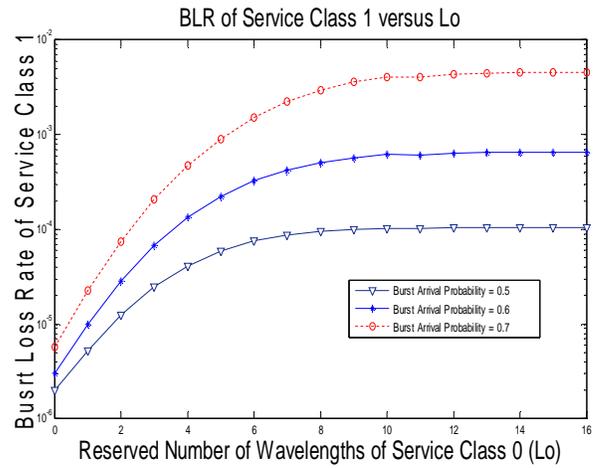

Fig. 14. The effect of reserved wavelengths (Lo) for service class 0 on burst loss rate of the other service class with $\rho$ = 0.5.

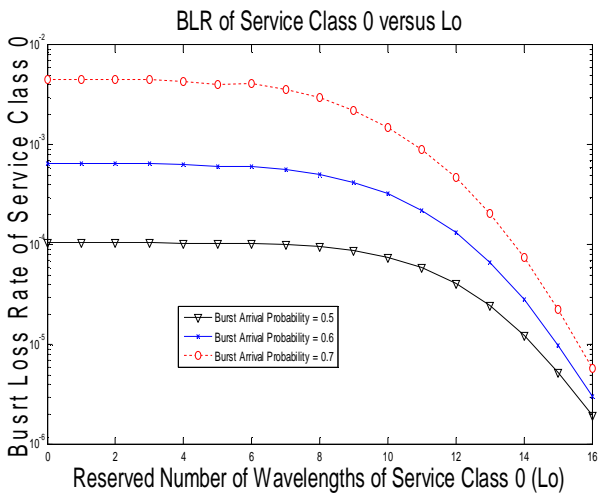

Fig. 12. The effect of reserved wavelengths (Lo) for service class 0 on burst loss rate of the same service class with $\rho$ = 0.5.

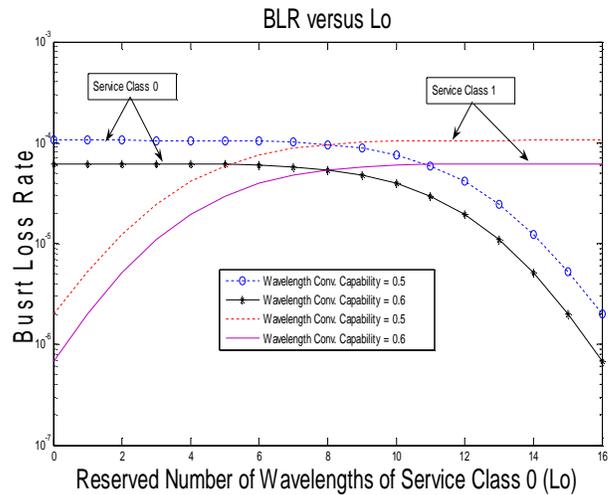

Fig. 15. The effect of reserved wavelengths (Lo) for service class 0 on burst loss rate of both the service classes with $A$ = 0.5.

# 6   CONCLUSION

This paper has presented an analytical model for burst loss rate of an optical burst switching network. The effects of several network design parameters on the system performance measures have been investigated and presented numerically. Results show that system performance improves with the increasing of number of channel wavelengths and wavelength conversion capabilities but degrades as the network traffic (burst arrival probability or number of slots per burst) or the burst blocking probability increases. From the result we also see that the entire burst is lost for unity blocking probability. The model results are found to be satisfactory agreement with the expected results.

**Md. Shamim Reza** was born in Magura, Bangladesh, on August 06, 1983. He received the B.Sc. and M.Sc. degrees in Electrical and Electronic Engineering from Bangladesh University of Engineering and Technology, Bangladesh, in 2006 and 2008, respectively. At present, he is an Assistant Professor with the Electrical and Electronic Engineering Department, Bangladesh University of Engineering and Technology, Dhaka, Bangladesh. His research interests include optical fibre communication, optical networking and signal processing.

**Md. Maruf Hossain** was born in Khulna, Bangladesh.. He received his B.Sc. Engg. degree in Electrical and Electronic Engineering from Bangladesh University of Engineering and Technology, Bangladesh, in 2006. Now he is doing his MSc Engg. in the Dept of Electrical and Electronic Engineering of Bangladesh University of Engineering and Technology, Bangladesh. At present, he is also a Lecturer with the Electrical and Electronic Engineering Department, American International University of Bangladesh, Dhaka, Bangladesh. His research interests include  optical fibre communication and wireless communications.

**Satya Prasad Majumder** is a Professor with the Electrical and Electronic Engineering Department, Bangladesh University of Engineering and Technology, Dhaka, Bangladesh. His research interests include  optical fibre communication, optical networking, mobile communications and signal processing.